# Improving the Equation of Exchange for Cryptoasset Valuation Using Empirical Data

Stylianos Kampakis, PhD, CStat, London Business School, Tesseract Academy[1]
Melody Yuan, MSc, University College London
Oritsebawo Paul Ikpobe, Tesseract Academy
Linas Stankevicius, Tesseract Academy

## Abstract

In the evolving domain of cryptocurrency markets, accurate token valuation remains a critical aspect influencing investment decisions and policy development. Whilst the prevailing equation of exchange pricing model offers a quantitative valuation approach based on the interplay between token price, transaction volume, supply, and either velocity or holding time, it exhibits intrinsic shortcomings. Specifically, the model may not consistently delineate the relationship between average token velocity and holding time. This paper aims to refine this equation, enhancing the depth of insight into token valuation methodologies.

---

[1] https://tesseract.academy

# Introduction

The valuation of cryptoassets is crucial for several reasons. Firstly, understanding the value of a cryptoasset is fundamental to making informed investment decisions. Given the high volatility [1, 2] and unique characteristics of cryptoassets, having a reliable valuation method can help investors assess the potential risks and returns associated with investing in a particular cryptoasset [3].

Secondly, regulators and financial institutions also need reliable and robust valuation models to assess the risk and potential impact of cryptoassets on the broader financial system [4]. Lastly, cryptoasset valuation is essential in designing the economics of a token, also known as tokenomics, which determines the long-term viability and sustainability of a crypto project [5]. Despite their increasing prominence, the valuation of cryptoassets remains a complex and contentious issue [6].

The primary research question addressed is: "How can empirical data refine the equation of exchange utilised in cryptoasset valuation?" A thorough analysis was undertaken on datasets sourced from CoinGecko, covering the period from 2014 to 2023. Utilising Python as the primary computational tool, this investigation delved into essential metrics such as the USD-denominated closing price, transaction volume, and market capitalisation. Advanced regression techniques were deployed to unearth and quantify the underlying dynamics of these metrics.

The findings introduce innovative equations that provide a reformed perspective on the relationship between velocity and holding time, resulting in enhanced equations of exchange pricing models. These refinements promise to augment the accuracy of token valuation, potentially setting new paradigms in cryptocurrency analytics.

# Background Information

## Equation of Exchange for the Valuation of Cryptoassets

Traditional valuation methods for financial assets, such as stocks and bonds, often rely on analyzing cash flows, earnings, or other financial metrics. These methods include discounted cash flow (DCF) models, discounted dividend models (DDM), price-to-earnings (P/E) ratios and book value calculations [7-9]. However, these traditional valuation methods may not be directly applicable to crypto assets.

Cryptoassets are not companies and do not have traditional financial statements to analyse [6]. While there exists real-asset-backed cryptoassets and security tokens [9, 10], cryptoassets generally do not generate earnings or cash flows in the same way that a company does, making it difficult to directly apply methods like DCF, P/E ratios or traditional price multiples [6, 10]

This has led to the exploration of alternative valuation methods, one of which is the application of the Quantity Theory of Money (QTM) by using the equation of exchange (EoE) for cryptoasset valuation [11-13], a quantitative valuation method. These applications posit that the value of a token can be determined by the token supply, token velocity, and transaction volume. Since some cryptoassets serve as a medium of exchange in their native token economies and also **in** the real world economy, usage of the EoE for valuation is valid [14].

The equation of exchange is defined as:

$$MV = PQ$$

Where :

*M* is the supply of money.
*V* is the velocity of money.
*P* is the average price level of goods and services.
*Q* is an index of the real value of aggregate expenditures.

However, it is important to note that the application of the EoE to cryptoassets is not without flaws. Not all cryptoassets can be considered as a medium of exchange [10, 15], reducing the validity of using the EoE for valuation. Some of the EoE models assume a stable velocity, which could be highly volatile for cryptoassets [16]. There is also literature debunking the assumptions made by some of the EoE models [17].

This paper aims to address these flaws by improving the equation of exchange based on empirical data, specifically focusing on velocity and holding time as a variable of the equation of exchange.

## Estimating velocity and holding time

Another issue with the EoE is that it is often difficult to identify reasonable values for the velocity. Conceptually, it is far easier for a tokenomist to use the holding time as a parameter, and define a prior or what a reasonable value is expected to be. The reason is that the holding time is directly related to staking incentives, or other mechanisms which incentivise a user to hold onto tokens instead of exchanging them [18, 19].

For this reason, Buterin proposed the inversion of the velocity as the definition of the holding time:

$$H = 1/V$$

However, this definition has been proven incorrect. While the quantity 1/V is related to the holding time, they are not equivalent [17].

In this study we use a linear regression model to estimate velocity through expected holding time to then use this estimate in a downstream model of price.

# Methods

## Dataset

For this research, historical data for eight key cryptoassets - BTC, BNB, CRV, DOT, ETH, LINK, UNI, and USDT - were used.

These assets were selected based on their classification within the cryptoasset taxonomy discussed in the literature review:

- BTC, ETH: Cryptocurrency
- BNB: Utility token, specifically exchange token
- CRV, UNI: Utility token, specifically DeFi Token
- DOT, ETH: Utility token, specifically smart contract platform token
- LINK: Utility token, specifically oracle token.
- USDT: Stablecoin

These assets are significant within their respective categories. BNB, CRV, DOT, ETH, LINK, and UNI primarily function as mediums of exchange within their ecosystems, while USDT serves as a store of value. BTC acts both as a general medium of exchange and a store of value. Including these diverse assets in our analysis allows for a more comprehensive analysis, addressing the limitations of the applicability of the existing equation of exchange (EoE).

The data was sourced from CoinGecko and downloaded in CSV format for ease of manipulation and analysis. CoinGecko was chosen for its extensive reach as the world's largest aggregator of historical cryptocurrency data, which is freely accessible and spans multiple exchanges and platforms [10, 20]. The dataset comprises several key metrics recorded daily at 00:00 UTC, from the inception of each crypto asset up to August 13, 2023.

## Experiment Methodology

The methodology is split into three parts:

1) Identify the best distribution fit for the velocity.
2) Creating a model of velocity derived from holding time.
3) Create a linear regression model that is using the derived velocity as an input.
4) Create a lookahead model that forecasts the price of a token one step ahead. Note that the timeseries followed a daily frequency.

Daily velocity is calculated as the ratio of the daily transaction volume to the market capitalisation. This is an average value across all the tokens in circulation. This metric indicates the number of times an average token is used in transactions in a day.

Given the lack of granular data on individual token velocities and holding times, an assumption is made that the inverse relationship between velocity and holding time applies at an aggregate level [21]. This constitutes a limitation of the study but is deemed acceptable within the scope of this study.

Holding time for each cryptoasset is subsequently calculated as the inverse of the daily velocity, which is equivalent to the ratio of market capitalisation to daily transaction volume. Note that this value is denominated in days. This is the amount of time an average token is held before being used for a transaction [22].

The performance metrics considered are Mean Absolute Error (MAE), Root Mean Squared Error (RMSE), and R-squared. Note that the lagged price EoE models are excluded from this evaluation. This is due to their inherent instability arising from their recursive nature making them unsuitable for time-series comparison [2].

# Experiment Steps

## Dataset preparation

The raw data has columns 'date', 'price', 'market_cap', and 'total_volume'.

The following steps were taken to preprocess the dataset:

1. Rename 'market_cap' to 'MC' and 'total_volume' to 'T'.
2. Calculate supply 'M' as the ratio of 'MC' to 'price', velocity 'V' as the ratio of 'T' to 'MC', and holding time 'H' as the ratio of 'MC' to 'T'.
3. Remove outliers by removing the bottom 10% of data in terms of velocity 'V'.
4. Then, calculate the derived velocity 'V'' using Equation 12, and the derived holding time 'H"' using Equation 11.

## Holding time estimation

The estimation of the holding time was split into two parts:

1) Identification of the general distributional form of the holding time.
2) Linear regression model

**Experimental steps: Distributional form identification**

1. Distribution Fitting with 'distfit

   Use distfit to identify the best fitting distribution for velocity and holding time.

2. Summary of Best-Fitting Distributions:

   The summary of the tested distributions and their RSS values are

   documented to identify any recurring patterns or trends in the test-fitting

   distributions across different cryptoassets

3. Iterative Testing Across Cryptoassets Repeat steps 1-3 for each and every cryptoasset.

**Experimental Steps: Linear regression**

1. Obtain log-normal distribution parameters

Obtain log-normal distribution parameters using the function on the daily velocity for each cryptoasset.

2. Generation of Synthetic Data

Assume a token economy consisting of 100 tokens. Assign each token a random velocity sample from a log-normal distribution with the parameters identified in Step 1. Calculate individual holding time by taking the inverse of velocity. Compute the mean velocity V and mean holding time H by averaging the individual token velocities and holding times. Repeat this process for a total of 1000 times to generate 1000 data points [23, 24].

3. Data Combination

Repeat Step 2 for each log-normal distribution, and then combine the data points from all the distributions except for USDT. This is because USDT is not a medium of exchange but rather a store of value as explained earlier [25, 26].

4. Outlier Removal using IQR method

Outliers are then removed from the combined data points using the interquartile range (IQR) method. The IQR is calculated as the difference between the 75"' percentile (Q3) and the 25"' percentile (Q1) of the data. Any data point that falls below Q1 - 1.5 * IQR or above Q3 1.5 * IQR is considered an outlier and is removed from the dataset to ensure robustness of the data used for regression. This was done to improve the model performance [27].

## EoE model estimation

We calculate the EoE Model price using linear regression with different polynomial terms and transformations.

The error was quantified through the use of adjusted R-squared.

# Results

# Estimation of distributional fit for the velocity and the holding time

Figure 1 below shows a histogram of the velocity of the estimated velocity times (logarithm and original values).

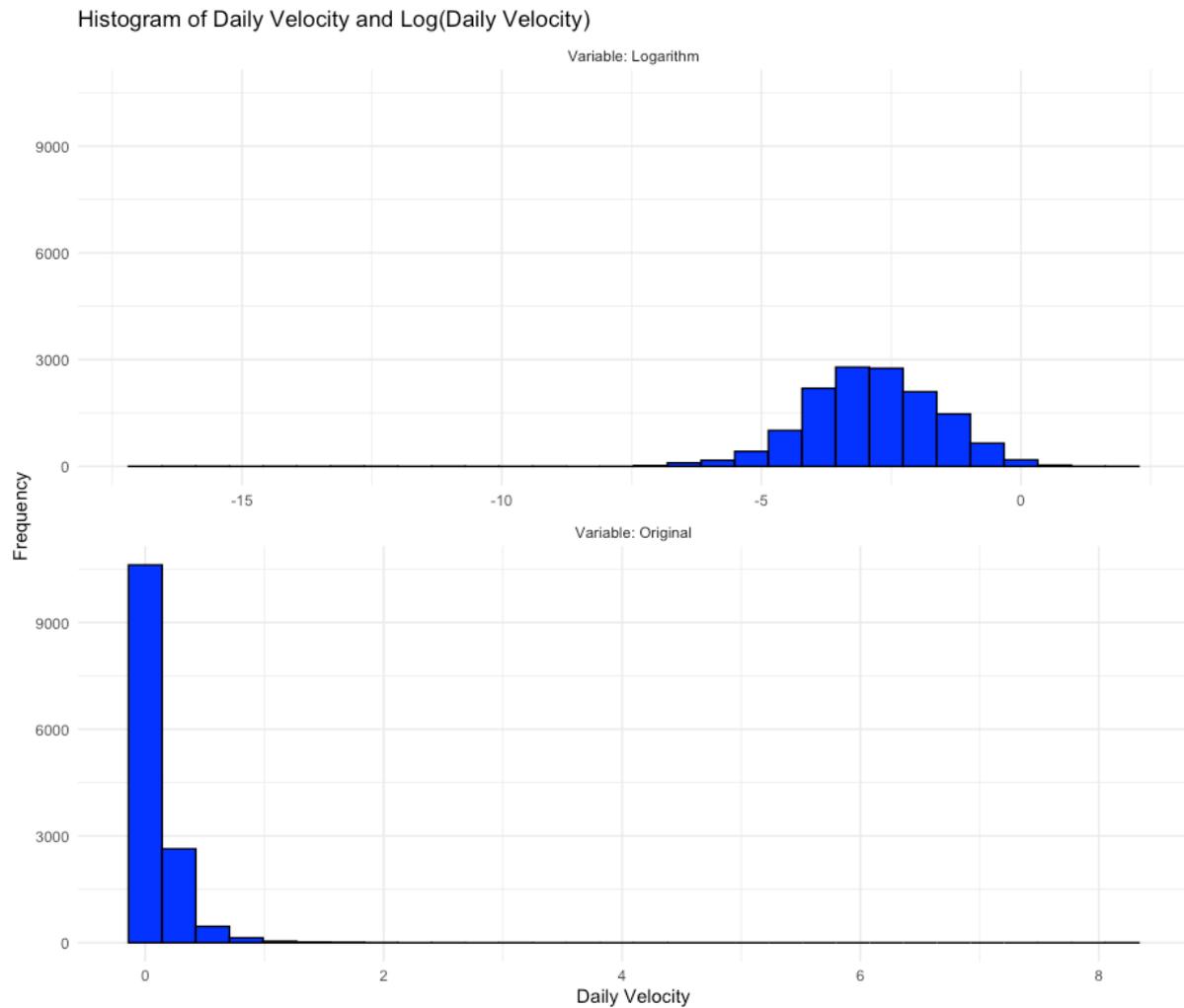

*Figure 1. Histogram of velocity*

Table 1 below shows the results of the distribution comparison.

*Table 1. Distribution comparison for the velocity. RSS stands for residual sum of squares.*

|      | BNB     | BTC     | CRV     | DOT    | ETH     | LINK   | UNI    | USDT   |
|------|---------|---------|---------|--------|---------|--------|--------|--------|
| Dist | lognorm | lognorm | lognorm | GEV    | lognorm | gamma  | GEV    | Pareto |
| RSS  | 52.087  | 5.716   | 0.334   | 33.455 | 4.394   | 3.052  | 38.087 | 0.029  |

Figure 2 shows the a histogram and a QQ-plot for the cryptoasset with the worst performance, which is BNB. It is evident that even in the worst possible case, the lognormal distribution provides a decent fit.

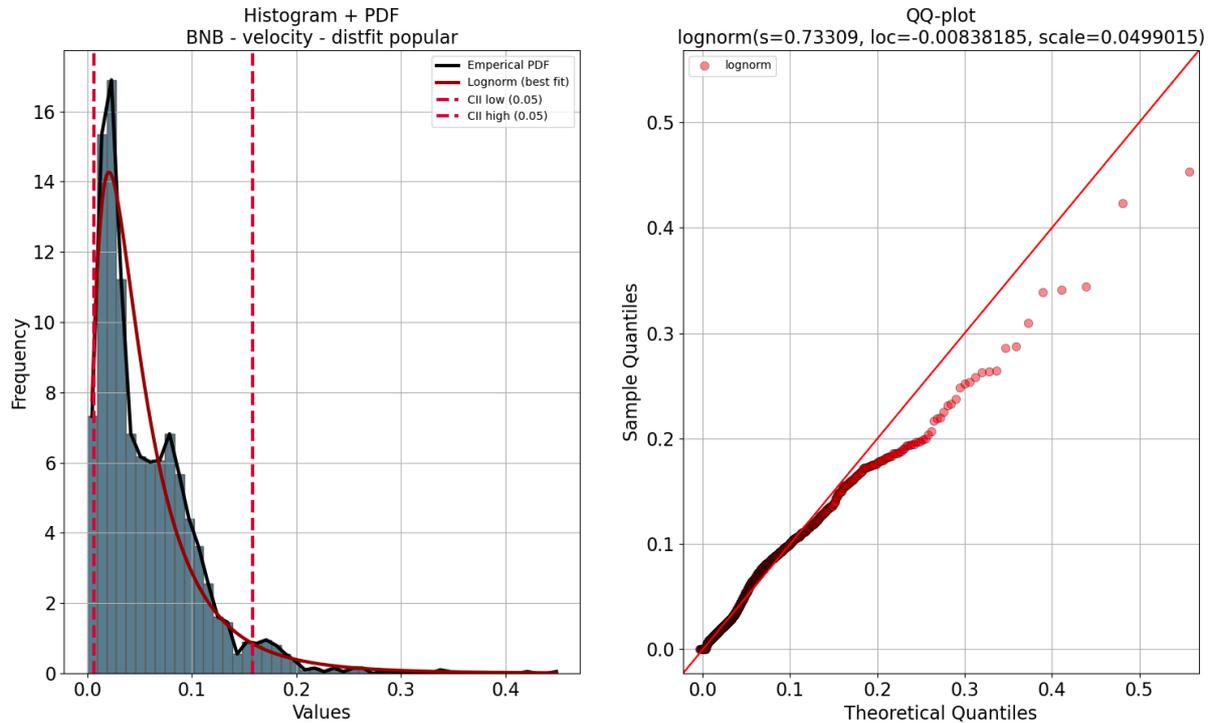

*Figure 2. Histogram and QQ-plot for the velocity of BNB*

Similar results can be drawn for the error of the lognormal distribution for the holding time.

*Table 2. Distribution fitting for the holding time*

|      | BNB      | BTC      | CRV      | DOT      | ETH      | LINK     | UNI      | USDT    |
|------|----------|----------|----------|----------|----------|----------|----------|---------|
| Dist | t        | GEV      | lognorm  | GEV      | lognorm  | GEV      | beta     | GEV     |
| RSS  | 0.000734 | 0.000037 | 0.000831 | 0.000047 | 0.000115 | 0.000216 | 0.000307 | 0.00252 |

It is evident that the log-normal distribution is the most frequently occurring best-fitting distribution for token velocity across various cryptoassets. Upon examining the summary of best-fitting distributions (detailed in the Appendix), it is noteworthy that for the cryptoassets where log-normal is not the top fit, it often ranks as the second or third best fit [28].

Moreover, the difference in RSS values between log-normal and the best-fitting distribution is generally small: 6.387 for DOT, 2.308 for LINK, and 0.036 for USDT. For UNI, although the difference in RSS is significant at 95.770, it is important to highlight that log-normal still ranks as the third best-fitting distribution [29].

This consistency in log-normal distribution as a strong fit across multiple cryptoassets suggests a common underlying mechanism influencing token velocity. It may indicate that most tokens in these markets have reached a level of stability in velocity, albeit with some outliers [30].

## Estimation of a linear regression model for velocity

A comparison across different model types was conducted to predict velocity from holding time.

All the model results are shown in Table 3 below, alongside the adjusted R^2.

The best model in this case is outlined in green.

*Table 3. Results of the velocity model against the holding time*

| Equation | Adj. R2 | Remarks | Valid Range |
|---|---|---|---|
| V'= 0.50420 + 0.00488 * H + 0.06262 * 1/H - 0.17128 * log(H) | 0.59427 | coef for 1/H has p-value > 0.05; non-monotonic relationship | H > 0 |
| V'= 0.67932 - 0.32288 * log(H) + 0.04242 * log(H)^2 | 0.59372 | non-monotonic relationship | H > 0 |
| V'= 0.46761 - 0.06727 * log(H) - 0.05569 * log(H)^2 + 0.01210 * log(H)^3 | 0.59372 | coef for log(H) has p-value > 0.05; non-monotonic relationship | H > 0.0665237 |
| V'= 0.03413 + 1.04035 * 1/H + 3.65049 * 1/H^2 - 14.52947 * 1/H^3 | 0.58948 | non-monotonic relationship | H > 2.32383 |
| V'= 0.02213 + 1.53561 * 1/H - 1.54978 * 1/H^2 | 0.58819 | non-monotonic relationship | H > 0.994961 |
| V'= 0.36493 - 0.02591 * H + 0.00075 * H^2 - 0.00001 * H^3 | 0.5877 | | H < 29.0523 |
| V'= 0.03358 + 1.20329 * 1/H | 0.58432 | | H > 0 |
| V'= 0.30099 - 0.01470 * H + 0.00022 * H^2 | 0.56667 | non-monotonic relationship | All H |
| V'= 0.37503 - 0.09028 * log(H) | 0.53936 | | 0 < H < 63.6931 |
| V'= 0.19854 - 0.00419 * H | 0.39442 | | H < 47.384 |

## Estimation of the equation of exchange

In this experiment, a variety of regression models were employed to rigorously investigate the relationship between average holding time H and average velocity V. The models were designed to capture linear, quadratic, cubic, and even logarithmic relationships between the variables. Note that the natural logarithmic transform was used in this experiment. Additionally, inverse transformations were applied to explore potential non-linear relationships. In the end, a log-linear model was found to provide satisfactory performance [31].

The model is defined in the equation below:

*log(price) = 0.88 log(T) + 0.84 log (M) + 1.15 log (1/V)*

The model achieves an adjusted R-squared of 0.97.

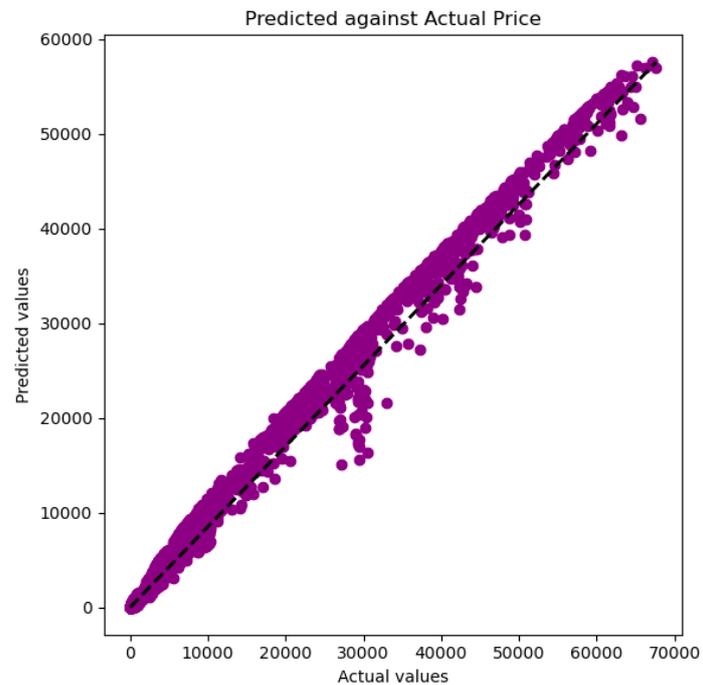

## Lookahead model

An extension of the model in the previous section was used with the addition of price at time t-1

$log(price)$ = 0.32 + 0.02 log($T$) + 0.04 log ($M$) + 0.03 log (V')
+ 0.98 log($price_{t-1}$)

The model was evaluated through the use of 20-fold cross-validation. It achieved an average Mean Absolute Error of 64.7 and RMSE of 97.19, with an $R^2$ of 0.93.

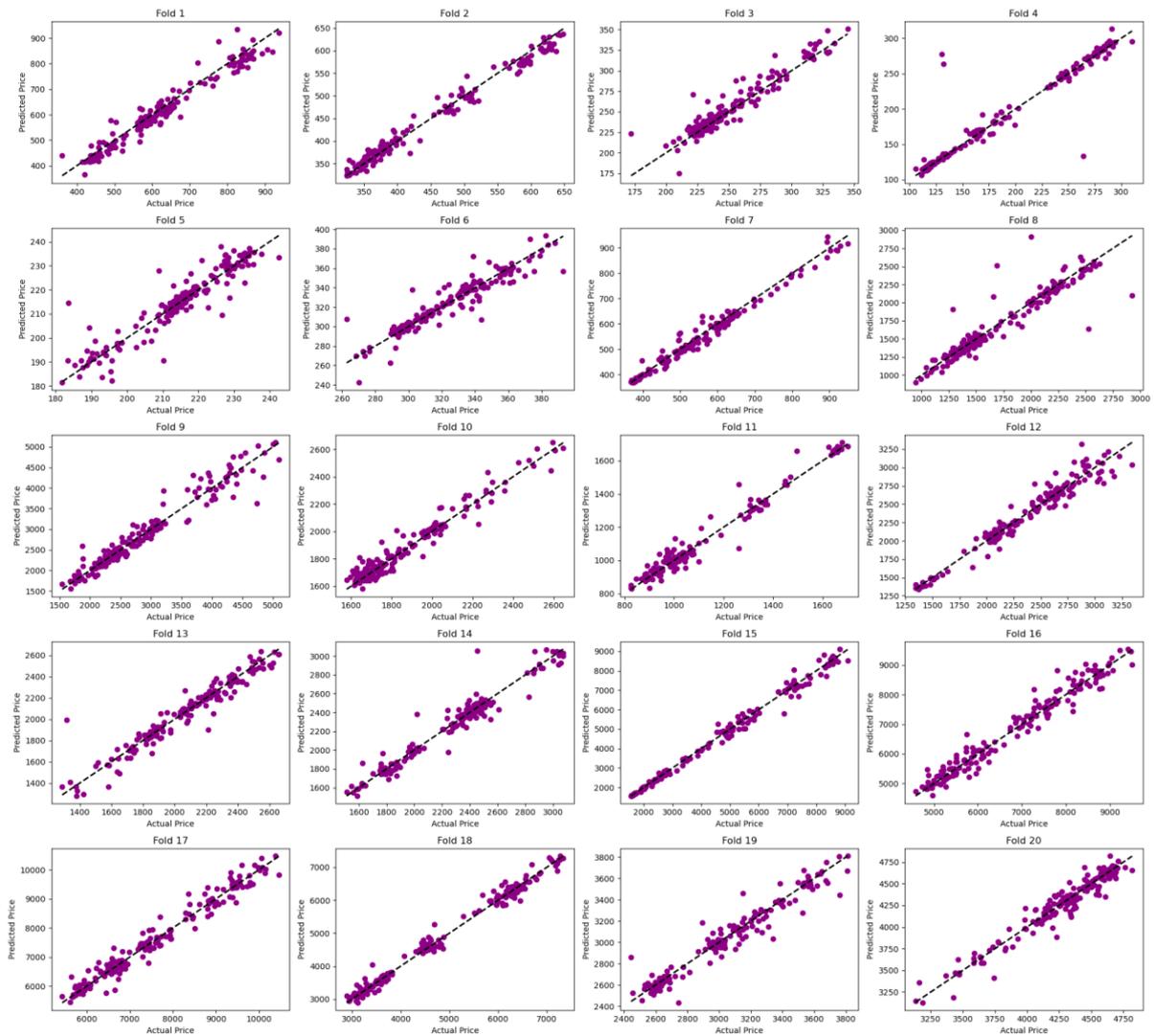

# Conclusion

Cryptoassets are a new class of financial assets. Traditional financial tools have limited applicability in valuing cryptoassets, warranting the need for other valuation tools. One prominent tool is the equation of exchange (EoE), of which token velocity and holding time have been identified to be key factors.

For this research, we verified and expanded on Scott Locklin's claim that mean holding time is not simply the inverse of mean velocity [35]. It was found that the distribution of velocity affects the mean holding time, even if the mean velocity was fixed at the same value. It was also found that empirical velocity and holding time generally follow a log-normal distribution across various cryptoassets. It was also found that a linear model with inverse transformation applied to holding time best described velocity.

We substituted this model for velocity and holding time into the baseline EoE models. A log-linear transformation was then applied to the baseline models, and regression was used to find the coefficients of the log-linear EoE models.

Lagged price was also added as a regression variable to this model. Both the log-linear and lagged price models achieved high adjusted R-squared values, indicating strong explanatory power.

**Limitations and Future Work**

However, there is still room for improvement. One limitation has been the lack of granular data on individual token velocities. Instead, aggregate values across all tokens were utilised. This does not capture as much fidelity, leading to reduced efficiency of the models developed using the aggregate values. Where possible, future work should use individual token velocities for analysis, perhaps using on-chain analysis tools as done by holden et al, 2019 [32], instead of the aggregate values [33].

Another limitation was the presence of some heteroscedasticity, autocorrelation, and non-normality of residuals observed **in** the regression experiments done in this dissertation. This reduces the accuracy and efficiency of the coefficients of the models found through Ordinary Least Squares regression, which was used to develop the equations quantifying the relationship between average holding time **H** and average holding time V, and the improved equation of exchange models [34, 35]. Weighted Least Squares should be explored as a method to quantify the relationship between **H** and V rather than Ordinary Least Squares regression [36].

Looking ahead, future studies might expand the scope of analysis to encompass a broader range and variety of medium of exchange tokens, thereby improving the generalisability of the valuation models. Additionally, concurrent qualitative analysis on top of the quantitative analysis done **in** this dissertation can be conducted to investigate the outliers in the data and performance of the models.